\documentstyle[prl,aps,twocolumn,epsfig]{revtex}

\begin{document}
\twocolumn[\hsize\textwidth\columnwidth\hsize\csname@twocolumnfalse\endcsname
\title{The shape of ecological networks}
\author{Michael L\"assig$^{(1)}$, Ugo Bastolla$^{(2)}$,
Susanna C. Manrubia$^{(2)}$, and Angelo Valleriani$^{(2)}$}
\address{$^{(1)}$Institut f\"ur theoretische Physik,
Universit\"at zu K\"oln, Z\"ulpicher Str. 77, 50937 K\"oln, Germany\\
$^{(2)}$Max-Planck-Institut f\"ur Kolloid- und
Grenzfl\"achenforschung, 14424 Potsdam, Germany}

\maketitle

\begin{abstract}

We study the statistics of ecosystems with a variable number of
co-evolving species. The species interact in two ways: by
prey-predator relationships and by direct competition with similar
kinds. The interaction coefficients change slowly through successful
adaptations and speciations. We treat them as quenched random
variables. These interactions determine long-term
topological features of the species network, which
are found to agree with those of biological systems.

\vspace{10 pt}

\vspace{24pt}
\end{abstract}

\vfill
]
\narrowtext

Population dynamics is a classical subject of evolutionary biology.
The mutually dependent dynamics of two or more populations or
species is often described by coupled differential equations
governing the relative change of the population sizes $N_i(t)$,
\begin{equation}
\frac{1}{N_i} \frac{{\rm d}N_i}{{\rm d}t} =
\sum_{j=1}^s g_{ij} N_j + h_i
\hspace{1cm} (i = 1,\dots, s) \; .
\label{LVgen}
\end{equation}
The interaction coefficients $g_{ij}$ can represent a prey-predator
relationship ($g_{ij} < 0$, $g_{ji} > 0$), direct competition
($g_{ij} < 0$, $g_{ji} < 0$), or mutualism ($g_{ij} > 0$,
$g_{ji} > 0$) between species $i$ and $j$, and the terms $h_i$
denote intrinsic production or death rates. These so-called
{\em Lotka-Volterra} equations, as well as many generalizations
thereof, have been used to model coexistence, invasions, and
adaptive change of populations. Of great importance is their
conceptual connection to mathematical {\em game theory} \cite{MS}.
A set of populations $N_1, \dots,N_S$ represents a mixed strategy.
For given interactions $g_{ij}$, an optimal strategy -- called
Nash equilibrium -- can often be realized as a stable fixed
point $N_1^*, \dots, N_S^*$ of an associated Lotka-Volterra dynamics.
This  explains how strategic optimization is reached in biological
systems through reproductive success, with no need for rational
thinking.

These equilibria determine the species' fate. For a given set of
equations (\ref{LVgen}), a species is viable if $N_i^* > 0$ and becomes
extinct if $N_i^* = 0$. Even the viable species are not perennial,
however. Successful adaptations, migrations, and
speciations (the splittings of a single species into a pair) eventually
change the number of players as well as the rules of the game, i.e., the
couplings $g_{ij}$. On large time scales,
this dynamics can be quite intermittent.
Correlated extinctions and  speciations alternate with periods of
relative stasis, leading to large fluctuations in the number
of species \cite{Nat}. Little is known on how this long-term behavior is
connected to the underlying interactions between species
in~(\ref{LVgen}).

\begin{figure}
\label{foodweb}
\centerline{\psfig{file=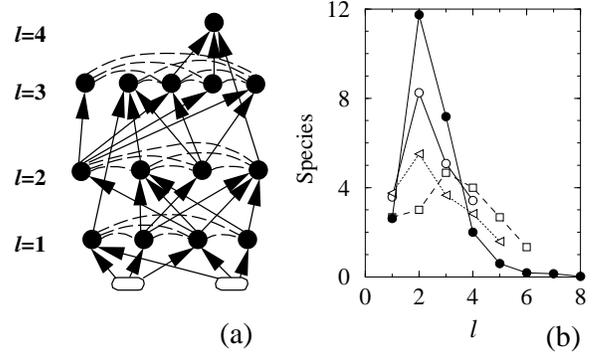,width=8cm}}
\caption{(a) The Pamlico estuary foodweb in North Carolina,
consisting of 14 species (filled circles) at four trophic levels.
Detritus, dinolagellates and diatoms are at the bottom
level ($l = 1$) and feed from external resources (empty symbols).
There is a single
trophic group at the highest level ($l = 4$), formed by the predatory
fishes {\it Roccus} and {\it Cynoscion}. Arrows point from prey
to predator; dashed lines connect species pairs with a nonzero
link overlap (see text). Data from Cohen (1990).
(b) Average species numbers for a set of natural ecosystems, taken 
from Cohen (1990) (empty symbols) and Rosenzweig (1995) (filled circles). This
last case corresponds to an average over 61 independent food webs, most
of which are empty at high levels.}
\end{figure}

The best studied natural ecosystems are {\em food webs}, i.e.,
communities of animal species in a closed environment where
food chains can be observed. Fig.~1(a) shows the graph of such a
network, each arrow representing a prey-predator relationship.
Despite large variations in size and environmental conditions,
large ecosystems share a few important topological characteristics:
(i)~Every species lives at a certain {\em trophic level}, which can
be defined as the minimum length of its relevant `downward' food
chains. Species at level one feed from external resources.
(ii)~The number of trophic levels is small, typically between three
and seven.
(iii)~Most species have a small number of relevant prey species
(typically around three), mainly from the next lower level.
(iv)~The number of species at level $l$ increases with $l$ for
lower values of $l$ and decreases again sharply for higher $l$
\cite{Cohen,webs}, see Fig.~1(b).
Networks of co-evolving species thus have a characteristic {\em shape}.

This remarkable structure calls for a theoretical explanation.
The classical work on Lotka-Volterra equations has established
stability
criteria for networks with random interactions
$g_{ij}$~\cite{Rmay}. In a real ecosystem, however,
the interactions are not random, but are themselves subject
to selection. Recently, Lotka-Volterra systems coupled to 
speciation \cite{Higgs} and immigration \cite{us} dynamics
have been studied by numerical simulations, and food web structures 
have indeed been found. Another class of
models focuses directly on the dynamics of extinctions and
speciations. These models have no explicit population dynamics
and mostly random topology, with the important exception of
Ref. \cite{Amaral}.

In this Letter, we present elements of a statistical theory for
large  ecosystems, using concepts and methods of theoretical physics.
We discuss the population dynamics with the {\em minimal} species
interactions consistent  with the observed complexity of ecosystems.
These interactions are  {\em  prey-predator relationships}, which
establish a flux of biomass between species, and {\em direct
competition} between similar species, which leads to their mutual
exclusion from ecological niches. The interaction coefficients
$g_{ij}$ in (\ref{LVgen}) are modeled as random variables that
change  through successful mutations. We focus on the (often
realistic) case that these mutations are sufficiently rare so that
the populations can reach stable equilibria in between.
In the language of statistical physics, the
species interactions are {\em quenched} random variables on the time
scales of population dynamics. The statistics of such ecosystems is
thus governed by a quenched distribution of Nash equilibria. This
distribution in turn emerges from a  long-term balance between
adaptations, speciations, and extinctions. The topology of the
resulting networks is found to be closely related to the underlying
dynamics of co-evolution. In the following, we concentrate on generic
topological features amenable to an approximate analytical treatment;
in particular, we derive the shape of ecosystems. A detailed
analysis of structure and dynamics of these networks will be published
elsewhere~\cite{long}.

To describe generic features of ecological networks, we choose
the simplest population dynamics containing predation and direct
competition. The interaction matrix in (\ref{LVgen}) is decomposed
accordingly, $g_{ij} = \gamma_{ij} - \beta_{ij}$. Predation
is parameterized by the constants $\gamma_{ij} = \gamma_+$ if $j$
is prey of $i$ and $\gamma_{ij} = - \gamma_-$ if $i$
is prey of $j$, with $0 < \gamma_+ < \gamma_-$.
Competition takes place for nesting places, mating opportunities,
and other resources not explicitly represented in the model.
It is strongest between individuals of the same species, but also
occurs between different species that interfere in each other's
livelihood~\cite{Gould}. We set $\beta_{ii} = 1$ (this normalization
amounts to an appropriate choice of the time scale in (\ref{LVgen}))
and $\beta_{ij} = \beta \rho_{ij}$ for $i \neq j$, with
$0 < \beta < 1$. The {\em link overlap} $\rho_{ij}$ measures
the degree of competition between the species. It is defined
as $\rho_{ij} \equiv c_{ij} / \sqrt{c_i c_j}$, where
$c_i$, $c_j$ are the number of predatory links of $i$, $j$, and
$c_{ij}$ is the number of common predatory links. (Species pairs
with a nonzero link overlap are connected by dashed lines in
Fig.~1.) Furthermore, all
species are assigned a uniform death rate $h_i = - \alpha$.
The external resources
are represented as a small number of extra 
`populations'  $N_i$ with $h_i = \gamma_+ R$ and 
predators only (i.e., $\gamma_{ij} < 0$ and $\beta_{ij} = 0$
for all $j$).

With these interactions, the fixed point populations given
by (\ref{LVgen}) can be written in the form
\begin{equation}
N_i^* = P_i - Q_i +h_i \;,
\label{Ni*}
\end{equation}
where $P_i = \gamma_+ \sum_{j \in \pi (i)} N_j^*
           - \gamma_- \sum_{j \in \Pi (i)} N_j^*$ is the
{\em productivity} of species $i$ from predation
(with $\pi(i)$ the set of its prey
and $\Pi(i)$ the set of its predators) and
$Q_i = \beta \sum_{j \ne i}
\rho_{ij} N_j^*$ is its
competition load.
Furthermore, we require a minimum population size $N_c \ll R$
for viable species, and count all species with $N_i^* < N_c$ as
extinct. Indeed, natural populations are known to be unstable under
short-term environmental fluctuations or adverse mutations if they
are too small or too dilute \cite{Allee}.

Of course, an ecosystem is not determined by its population
dynamics alone but also by the long-term processes of successful
mutations, in particular, speciations \cite{Higgs,Leibold}.
In this model, a mutation is represented as a
stochastic change of predation links that is consistent
with existing food chains. It turns out that details of
this process are not relevant for our present purpose of
deriving global network characteristics. It is sufficient
to assume that speciations and adaptations maintain a broad
distribution of productivities $P_i$, and hence, of population
sizes $N_i^*$ (in a sense made precise below).
This is well supported by field observations
and by our numerics~\cite{long,Pielou}. An increase in the number
of species reduces the average productivity and increases
the average competition load. Hence, such an ecosystem
admits only a certain number of viable species, whose
productivities satisfy $P_i > Q_i + \alpha + N_c$.
The number of these {\em ecological niches} depends on
the interaction parameters $\beta$, $\gamma_+$, $\gamma_-$, and
on the dimensionless ratios $R/N_c$, $\alpha/N_c$.
Once the niches are filled, ongoing speciations and the
subsequent adaptations reshuffle the productivities and the population
numbers $N_i^*$ of all the species, forcing  the least viable
ones into extinction. On large time scales, this is a stationary
stochastic process. The relative success of an individual species keeps
changing as a result of its own adaptations and those of the other
species, resulting in a constant threat of extinction called
the {\em Red Queen} effect~\cite{Stenseth}.
The shape of these mature networks is determined essentially
by the distribution of ecological
niches. To see this, consider
first two cases of simple networks with fixed topology.

{\em 1. A single food chain}
is a community of $L$ species on $L$ trophic levels. The species
at level one feeds from an external resource, the species at level
$l$ from that at level $l-1$ ($l = 2,3,\dots,L$). The productivities
of this chain are given by the equations
\begin{equation}
P_l = \gamma_+ N^*_{l-1} - \gamma_- N^*_{l+1}
      \hspace{1cm} (l = 1,\dots,L)
\label{Pl}
\end{equation}
and $P_0 = -\gamma_1 N_1^*$, with the boundary condition $N^*_{L+1} = 0$. 
They determine directly the population numbers
$N_l^* = P_l - \alpha$ since all competition loads vanish.  The
entire chain is viable if $P_l > P_c$ for all species $l$, with
the minimum productivity
\begin{equation}
P_c = \alpha + N_c\;.
\label{Pc}
\end{equation}
The equations (\ref{Pl}) can be solved exactly by recursion
starting from the top level $l = L$.  For the biologically
important case of small $\gamma_+$, we find that the maximum value
of $L$ compatible with (\ref{Pl}) is
\begin{equation}
L = \frac{-1}{\log \gamma_+} \,
    \log \! \left (\frac{R}{{\rm const.} \, \alpha + N_c} \right )
    -1 + O(\gamma_+)
\end{equation}
by applying the condition (\ref{Pc}) at the top level.

The parameters
$\alpha$ and $N_c$ are seen to be equivalent viability cutoffs for
the chain since they reduce primarily the top population $N_L^*$.
More generally, the population numbers
$N_l^*$ are found to be rapidly decreasing with increasing $l$ for
all relevant parameter values.  Hence, as observed in nature,
viable chains are always short~\cite{Rosen}.

{\em 2. A single trophic level} is a group of $S$ species that may
have a significant overlap in their predation links and a resulting
competition load. First we consider the productivities $P_i$ as fixed
by the interactions with other trophic levels and concentrate on
the effects of the direct competition terms $Q_i$. In a
`mean field' approximation, we replace the individual link overlaps
by an expectation value $\bar \rho$ depending on the predation
clusters. In the simplest case of  random
predation, Eq.~(\ref{Ni*}) then determines the
fixed point populations
\begin{equation}
N_i^* =
 \frac{P_i - \beta \bar \rho S \bar N - \alpha}{1 - \beta \bar
 \rho}\;;
\end{equation}
the average
$\bar N \equiv S^{-1} \sum_{i=1}^S N_i^*$ is given by
\begin{equation}
\bar N = \frac{\bar P - \alpha}{ 1 + \beta \bar \rho (S-1)} \;.
\label{Nbar}
\end{equation}
The viability of all species ($N^*_i > N_c$) again sets a
minimum productivity
\begin{equation}
P_c = \alpha + (1 - \beta \bar \rho) N_c  + \beta \bar \rho S \bar N \;.
\label{Pc2}
\end{equation}
We now use the assumption that the productivities $P_i$ are
drawn from a broad probability distribution given by
$\Phi (q) \equiv {\rm Prob}(P_i/\bar P < q)$.
(The qualitative results do not depend strongly on the form of
$\Phi (q)$; here we use a simple approximation~\cite{Phi}.)
 The species
community becomes unstable if the least viable species has
a productivity below $P_c$.
The number of species in a mature trophic level
can therefore be estimated from
the relation $S \Phi(P_c/\bar P) = O(1)$. Eq.~(\ref{Pc2}) then
becomes an implicit relation for $S$ as a function of
$\bar P/N_c$, $\alpha / N_c$, and the average pairwise competition
load $\beta \bar \rho$.
Consider, for example, a trophic level with random predation from
a set of $S'$ prey species from the levels below. Using a simple
approximation for the average link overlap
$\bar \rho (S,S')$~\cite{overlap},
it can be shown that the solution
of Eq.~(\ref{Pc2}) always satisfies
$S \leq \max( a(\beta) S'/ {\bar c}, 1)$,
where $a(\beta) \ge 1$ and 
$\bar c$ is the average number of prey species per predator species.
That is, competition determines the number of ecological niches in a
trophic level as a function of the prey diversity and the
competition strength $\beta$. For sufficiently large $\beta$,
only non-overlapping
species can coexist, i.e., $S = \max(S'/{\bar c}, 1)$.
This result generalizes the well
known theorem of competitive exclusion \cite{ecology}, which states
the condition for coexistence of two competing species. Note that
this limiting effect on the number of species exists independently of
the population numbers. It is indeed crucial for the buildup
of high population numbers at the lower trophic levels.
For example, a trophic level feeding from effective resources of size
$R \gg \alpha, N_c$ 
acquires an extensive population number per species  
 $\bar N \sim R/S$, while $S$ is asymptotically
independent of $R$.  Without competitive 
exclusion ($\beta = 0$), speciations would
further increase $S$. This leads eventually to an extensive 
number of marginally viable species, i.e., $S \sim R/\bar N$
with  $\bar N$ of
order $N_c$. Such a level could not support sizeable predation from above.

We now turn to a full ecological network with $L$ trophic levels.
In the mean field approximation, we treat all species at the same
level on an equal footing and derive self-consistent equations
for the level averages of  population and species number,
$\bar N_l$ and $S_l$ ($ l = 1,\dots,L$).
The average productivities $\bar P_l$ satisfy
the recursion relations
\begin{equation}
\bar P_l = \gamma_+ {\bar c} \bar N_{l-1} -
           \gamma_- {\bar c} (S_{l+1}/S_l) \bar N_{l+1} \;,
\label{Pl2}
\end{equation}
where we assume that the species at every level predate randomly
on the species at the next lower level.
The average number $\bar c$ of predation links per predator 
is taken to be independent of $l$;
this is indeed suggested by field data. The average number of predators
per prey is then simply ${\bar c} S_{l+1}/S_l$. The productivity
$\bar P_l$ is linked to $\bar N_l$ and $S_l$ as in (\ref{Nbar}),
using for $\bar \rho(S_l, S_{l-1})$ the same approximation as
above~\cite{overlap}. Hence, the relations (\ref{Pl2}) determine
the population numbers given the species numbers. The latter
are again limited by the stability criteria
$S_l \Phi (P_{c,l} / \bar P_l) = O(1)$ with the minimum productivities
$P_{c,l}$ given as in (\ref{Pc2}); these relations determine the $S_l$
given the $\bar N_l$. The coupled set of equations
can be solved iteratively. Finally, the number of levels $L$ follows
from the condition $\bar N_L \approx N_c$, which is equivalent
to $S_L \approx 1$.

Over a wide range of relevant parameters, these networks have
the characteristic shape shown in the example of Fig.~2: The species
numbers $S_l$ increase with $l$ at low levels due to the increasing
prey diversity, which opens up more and more niches. They reach a
maximum at an intermediate level and decrease again at higher levels,
because more and more species have population numbers too low to
support further predation. Hence, these two regimes reflect
the two kinds of species interactions. The population numbers show an
approximately exponential decrease in both regimes, just like for
a single vertical chain. Hence, $L$ is always small, in agreement with
observations and with the results of \cite{Higgs,us}.
The functional form of the patterns $S_l, \bar N_l$ can
be described by analytical approximations depending on
the parameter values.

\begin{figure}
\label{Nc}
\centerline{\psfig{file=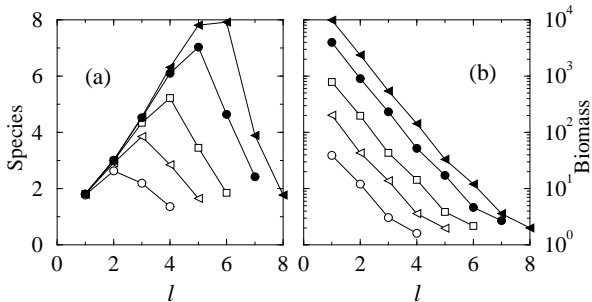,width=8cm}}
\caption{The shape of ecosystems. (a) The species numbers
$S_l$ ($1 \leq l \leq L$) for networks with $L$ trophic levels.
The parameters are
$\bar c=3$,
$\gamma_+ = 0.3$,
$\gamma_- = 2.0$, $
\beta = 0.2$,
$q_0=0.35$ (see [17]),
$\alpha/N_c = 1 $, and
$R/N_c = 2 \times 10^3, \; 10^4, \; 4 \times 10^4, \; 2 \times 10^5,$ and
$5 \times 10^5$ for the cases $L = 4, \; 5, \; 6, \;7$, and $8$,
respectively. (b) The average population numbers $\bar N_l$ for the
same cases as in (a).}
\end{figure}

Species networks are thus quite far from randomly connected.
Their topological shape is dynamically
generated by the coupled evolution
of populations and the slower adaptative changes. The ubiquity
of this shape suggests
that predation and competition of similar species are the
fundamental interactions governing the long-term coevolution
of large ecosystems. They are remarkably simple. Predation
is the basic transport of energy in the system, competition
forces the species into states with little overlap. In physics,
mutual avoidance is a well known property of fermions. Competitive
exclusion may thus be regarded as the Pauli principle of co-evolution:
It generates the complexity of species networks just as its
quantum-mechanical counterpart does for atoms and molecules.

We have discussed here the {\em global} shape of these networks. It
remains a challenging task to explore the connection between dynamics
and topology {\em locally}, that is, at the level of individual species and
their genealogies.

M.L. is grateful to the MPI for Colloids and Interfaces for the
kind hospitality throughout this work.


\begin{thebibliography}{99}
\bibitem{Rosen} The relevance of energetic limitations for the length of
real food chains is a debated issue (see for example \cite{Ros}).
\bibitem{Phi} This approximation is $\Phi (q) = (q - q_0)/2(1-q_0)$
with $0 < q_0 \ll 1$,
which gives $S = (1 - q_0)/((P_c/\bar P) - q_0)$.
\bibitem{overlap}
For $S \le \max(S'/{\bar c},1)$, a configuration with zero overlap exists.
For larger values of $S$, we assume random predation. This gives
$$ \bar \rho (S,S') =
  \left \{ \begin{array}{ll}
           0                & \mbox{if } S \leq \max(S'/\bar c,1) \\
	   \min(\bar c/S',1)& \mbox{if } S > \max(S'/\bar c,1)
	   \end{array}
  \right. \; . $$
Other approximations of the predation clusters are possible.
\end{thebibliography}
\end{document}